\documentclass[traditabstract,twocolumn]{aa}

\usepackage{graphicx}
\usepackage{epsfig}
\usepackage{amssymb}
\usepackage{natbib}
\usepackage{txfonts}
\usepackage{pslatex}
\usepackage{caption}
\usepackage{dcolumn}% Align table columns on decimal point
\usepackage{bm}% bold math
\usepackage{subfig}

\newcommand\simlt{\lower.5ex\hbox{$\; \buildrel < \over \sim \;$}}

\begin{document}

\title{The high-redshift star formation rate derived from GRBs: possible origin and cosmic reionization}

\author{F. Y. Wang\inst{1,2,3}}

\institute{ School of Astronomy and Space Science, Nanjing
University, Nanjing 210093, China; \and Department of Physics, The
University of Hong Kong, Pokfulam Road, Hong Kong, China; \and Key
Laboratory of Modern Astronomy and Astrophysics (Nanjing
University), Ministry of Education, Nanjing 210093, China; }

\authorrunning{Wang}
\titlerunning{The high-redshift SFR derived from GRBs and cosmic reionization}
\begin{abstract}
{The collapsar model of long gamma-ray bursts (GRBs) indicates that
they may trace the star formation history. So long GRBs may be a
useful tool of measuring the high-redshift star formation rate
(SFR). The collapsar model explains GRB formation via the collapse
of a rapidly rotating massive star with $M> 30M_\odot$ into a black
hole, which may imply a decrease of SFR at high redshift. However,
we find that the \emph{Swift} GRBs during 2005-2012 are biased
tracing the SFR, including a factor about $(1+z)^{0.5}$, which is in
agreement with recent results. After taking this factor, the SFR
derived from GRBs does not show steep drop up to $z\sim 9.4$. We
consider the GRBs produced by rapidly rotating metal-poor stars with
low masses to explain the high-redshift GRB rate excess. The
chemically homogeneous evolution scenario (CHES) of rapidly rotating
stars with mass larger than $12M_\odot$ is recognized as a promising
path towards collapsars in connection with long GRBs. Our results
indicate that the stars in the mass range $12M_\odot<M<30M_\odot$
for low enough metallicity $Z\leq 0.004$ with the GRB efficiency
factor $10^{-5}$ can fit the derived SFR with good accuracy.
Combining these two factors, we find that the conversion efficiency
from massive stars to GRBs is enhanced by a factor of 10, which may
be able to explain the excess of the high-redshift GRB rate. We also
investigate the cosmic reionization history using the derived SFR.
The GRB-inferred SFR would be sufficient to maintain cosmic
reionization over $6<z<10$ and reproduce the observed optical depth
of Thomson scattering to the cosmic microwave background.}
\end{abstract}


\maketitle

\section{Introduction}
Gamma-ray bursts (GRBs) are the brightest electromagnetic explosions
in the universe (for a recent review, see Gehrels et al. 2009).
Because of their very high luminosity, GRBs can be detected out to
the edge of the visible Universe (Ciardi \& Loeb 2000; Lamb \&
Reichart 2000; Bromm \& Loeb 2002, 2006). The farthest GRB to date
is GRB 090429B with a photometric redshift $z=9.4$ (Cucchiara et al.
2011), significantly larger than those of the most distant quasars.
This property makes GRBs indispensable beacons to study the early
universe, including the star formation rate (Totani 1997; Wijers et
al. 1998; Porciani \& Madau 2001; Bromm \& Loeb 2002,2006), the
intergalactic medium (IGM) (Barkana \& Loeb 2004; Inoue et al. 2007;
McQuinn et al. 2008), and the metal enrichment history (Savaglio
2006; Wang et al. 2012). In addition, GRBs have been used as
standard candles to constrain cosmological parameters and dark
energy (Dai, Liang \& Xu 2004; Schaefer 2007; Wang, Qi \& Dai 2011).

The most popular theoretical model of long-duration GRBs is the
collapse of a massive star to a black hole (Woosley 1993).
Observations also show that GRBs are associated with Type Ib/c
supernovae (Stanek et al. 2003; Hjorth et al. 2003). So GRBs provide
a complementary technique for measuring the SFR history (Totani
1997; Wijers et al. 1998; Porciani \& Madau 2001). Recent studies
show that \emph{Swift} GRBs are not tracing the star formation
history measured by traditional means exactly but including an
additional evolution (Le \& Dermer 2007; Salvaterra \& Chincarini
2007; Kistler et al. 2008; Y\"{u}ksel et al. 2008; Wang \& Dai 2009;
Wanderman,\& Piran 2010; Qin et al. 2010; Cao et al. 2011; Robertson
\& Ellis 2012; but see Elliott et al. 2012). The SFR inferred from
the high-redshift ($z>6$) GRBs seems to be too high in comparison
with the one obtained from some high-redshift galaxy surveys
(Kristler et al. 2009; Bouwens et al. 2009). Kistler et al. (2008)
found that there are about four times as many GRBs at redshift
$z\sim 4$ than expected from star formation measurements. They
claimed that some unknown mechanism is leading to an enhancement
about $(1+z)^{\delta}$ ($\delta=1.5$) in the observed rate of
high-redshift GRBs. Using more \emph{Swift} data, Kistler et al.
(2009) found a slightly lower value of enhancement about
$(1+z)^{1.2}$. Robertson \& Ellis (2012) found the value of $\delta$
is about $0.5$ by comparing the cumulative redshift distribution of
GRBs and SFR. But on the other hand, Elliott et al. (2012) found
that the value of $\delta$ is about zero using a small sample of
GRBs. In order to explain this discrepancy, many models have been
proposed. Li (2008) explained the observed discrepancy between the
GRB rate history and the star formation rate history as being due to
cosmic metallicity evolution (Langer \& Norman 2006), by assuming
that long GRBs tend to occur in galaxies with low metallicities.
Cheng et al. (2010) suggested that this discrepancy could be solved
if some high-redshift GRBs are produced by superconducting cosmic
strings. Wang \& Dai (2011) used an evolving initial mass function
(IMF) of stars to explain the GRB redshift distribution. Virgili et
al. (2011) discussed the possibility that the evolution of the GRB
luminosity function break with redshift may explain this
discrepancy. Observations also show differences in the population of
GRB host galaxies compared to expectations for an unbiased
star-formation tracer (Tanvir et al. 2004; Fruchter et al. 2006;
Svensson et al. 2010).

In this paper, we study the star formation rate history derived from
GRBs. First we use the \emph{Swift} GRB sample to test the evolution
of GRB rate relative to SFR. If GRBs trace star formation in the
universe without bias, the ratio of the GRB rate to the SFR would
not be expected to vary with redshift. We find that this ratio is
proportional to $(1+z)^{0.5}$. The index is smaller than the value
of Kistler et al. (2009). We also derive the high-redshift SFR using
\emph{Swift} GRB sample by correcting this evolution. Then, we
consider the rapidly rotating metal-poor stars with masses smaller
than critical mass $M_{\rm cri}\sim 30 M_\odot$ to see if they can
produce GRBs to explain the discrepancy between high-redshift SFR
and GRB rate. The collapsar model indicates that stars with mass
larger than $30M_{\odot}$ can produce GRBs (Woosley 1993; Bissaldi
et al. 2007; Raskin et al. 2008). Observation also shows that the
progenitor of GRB 060505 has a mass above $30M_\odot$ (Th\"{o}ne et
al. 2008). Yoon \& Langer (2005) investigated the evolution of
rotating single stars in the mass range $12M_\odot<M<60M_\odot$ at
low metallicity. They found that if the initial spin rate is high
enough, the time scale for rotationally induced mixing becomes
shorter than the nuclear time scale. The star may evolve in a
quasi-chemically homogeneous way. In particular, for low enough
metallicity, this type of evolution can lead to retention of
sufficient angular momentum in cores to produce GRBs according to
the collapsar scenario. Last, we calculate the impact of this
GRB-inferred star formation rate on the reionization history,
including the optical depth of electron scattering to the cosmic
microwave background.

The structure of this paper is arranged as follows. In the next
section, we compile the \emph{Swift} GRB sample till GRB 110403 and
test the evolution of GRB rate. The SFR derived from GRBs is given
in section 3. We show the model of GRBs from chemically homogeneous
evolution scenario and the influence on high-redshift SFR derived
from GRBs in section 4. We compute the reionization history with
this GRB-inferred SFR in section 5. We conclude with a summary in
section 6.
\section{The latest GRB sample}

\begin{figure}
\centering
\includegraphics[width=0.5\textwidth]{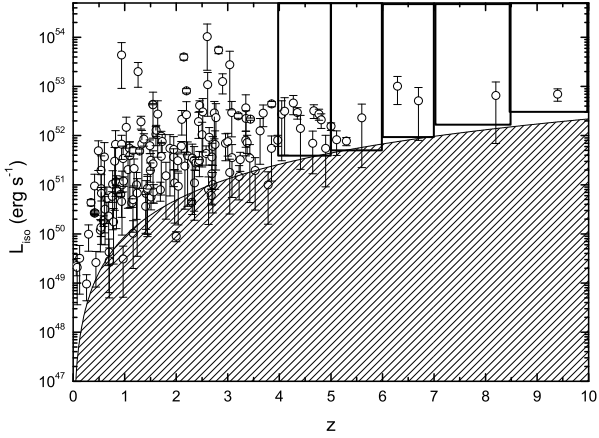} \caption{Distribution of the
isotropic-equivalent luminosity for 157 long-duration \emph{Swift}
GRBs. Demarcated are the GRB subsamples used to estimate the SFR.
The shaded area approximates the detection threshold of \emph{Swift}
BAT.} \label{GRBnum}
\end{figure}

The expected redshift distribution of GRBs is
\begin{equation}
\frac{d N}{d z}=F(z) \frac{\varepsilon(z)\dot{\rho}_*(z)}{\langle
f_{\rm beam}\rangle} \frac{dV_{\rm com}/dz}{1+z},
\end{equation}
where $F(z)$ represents the ability both to detect the trigger of
burst and to obtain the redshift, $\varepsilon(z)$ accounts for the
fraction of stars producing GRBs, $\dot{\rho}_*(z)$ is the SFR
density. The $F(z)$ can be treated as constant when we consider the
bright bursts with luminosities sufficient to be detected within an
entire redshift range, so $F(z)=F_0$. GRBs that are unobservable due
to beaming are accounted for through $\langle f_{\rm beam}\rangle$.
The $\varepsilon(z)$ can be parameterized as
$\varepsilon(z)=\varepsilon_0(1+z)^\delta$, where $\varepsilon_0$ is
an unknown constant that includes the absolute conversion from the
SFR to the GRB rate in a given GRB luminosity range. Kistler et al.
(2008) found the index $\delta=1.5$ from 63 \emph{Swift} GRBs. A
little smaller value $\delta=1.2$ was inferred using 119
\emph{Swift} GRBs (Kistler et al. 2009). In a flat universe, the
comoving volume is calculated by
\begin{eqnarray}
    \frac{dV_{\rm com}}{d z} = 4\pi D_{\rm com}^2 \frac{dD_{\rm com}}{d z} \;,
\end{eqnarray}
where the comoving distance is
\begin{eqnarray}\label{com}
    D_{\rm com}(z) \equiv \frac{c}{H_0} \int_0^z \frac{dz^\prime}{\sqrt{
            \Omega_m(1+z^\prime)^3 + \Omega_\Lambda}} \;.
\end{eqnarray}
In the calculations, we use $\Omega_m=0.27$, $\Omega_\Lambda=0.73$
and $H_0$=71 km~s$^{-1}$~Mpc$^{-1}$ from the \emph{Wilkinson
Microwave Anisotropy Probe} (WMAP) seven-year data (Komatsu et al.
2011).

We use the latest \emph{Swift} long-duration GRB sample till GRB
110403. The data is taken from Butler et al. (2007,2010) and
website\footnote{$\rm http://astro.berkeley.edu/\sim
nat/Swift/bat\_spec\_table.html$}. The isotropic-equivalent
luminosity of a GRB can be obtained by
\begin{equation}
L_{\rm iso}=E_{\rm iso}(1+z)/T_{90}.
\end{equation}
The distribution of $L_{\rm iso}$ for 157 GRBs in the sample is
shown in Figure~\ref{GRBnum}. We use the same luminosity cuts in
these redshift bins as Kistler et al. (2009). The shaded area
approximates the detection threshold of \emph{Swift} BAT, which can
be calculated as follows. The luminosity threshold can be
approximated by a bolometric energy flux limit $F_{\lim} = 1.2
\times 10^{-8}$erg cm$^{-2}$ s$^{-1}$. The luminosity threshold is
then
\begin{eqnarray}
    L_{\lim} = 4\pi D_L^2 F_{\rm lim} \;,   \label{lum_lim}
\end{eqnarray}
where $D_L$ is the luminosity distance to the burst.

\begin{figure}
\centering
\includegraphics[width=0.5\textwidth]{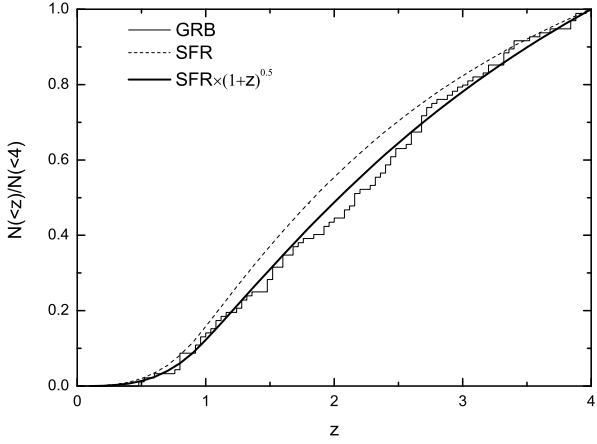} \caption{Cumulative distribution of 92 \emph{Swift} GRBs with
$L_{\rm iso}>10^{51}\rm erg~s^{-1}$ in $z=0-4$ (stepwise solid
line). The dashed line shows the GRB rate inferred from the star
formation history of Hopkins \& Beacom (2006). The solid line shows
the GRB rate inferred from the star formation history including
$(1+z)^{0.5}$ evolution.} \label{cum}
\end{figure}

In order to test the GRB rate relative to the SFR, we must choose
bursts with high luminosities, because only bright bursts can be
seen at low and high-redshifts, so we choose the luminosity cut
$L_{\rm iso}> 10^{51}$ erg s$^{-1}$ (Y\"{u}ksel et al. 2008) in the
redshift bin $0-4$. This removes many low-redshift, low-$L_{\rm
iso}$ bursts that could not have been seen at higher redshift.
Because the SFR at high redshift is poorly known (Bouwens et al.
2012; Oesch et al. 2013; Coe et al. 2013), so we choose the redshift
range $0<z<4$, in which SFR is well measured. We have 92 GRBs in
this subsample. We use the SFR history from Hopkins \& Beacom
(2006). We compare the predicted and observed cumulative GRB
distributions in Figure~\ref{cum}. We find that the
Kolmogorov-Smirnov statistic is minimized for $\delta=0.5$, which is
consistent with Robertson \& Ellis (2012). At the $2\sigma$
confidence level, the value of $\delta$ is in the range
$-0.15<\delta<1.6$. Our result is smaller than the values in Kistler
et al. (2009). This may be account for the different GRB sample. The
GRB sample observed by Swift during 2005-2012 is used in this paper.
This bias must be taken into account when one relates the GRB rate
to the SFR.

\section{The SFR derived from GRBs}

\begin{figure}
\centering
\includegraphics[width=0.5\textwidth]{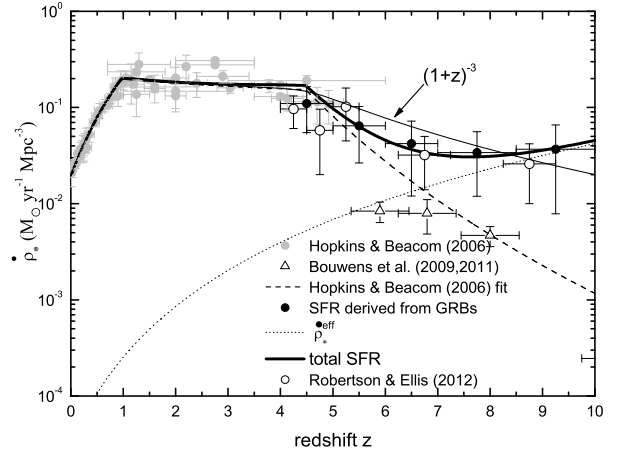} \caption{The cosmic star formation history.
The grey points are taken from Hopkins \& Beacom (2006), the dashed
line shows their fitting result. The triangular points are from
Bouwens et al. (2009, 2011). The open circles are taken from
Robertson \& Ellis (2012). The filled circles are the SFR derived
from GRBs in this work. The dotted line shows the effect SFR
calculated from equation (\ref{effsfr}). The thick solid line shows
the combination of dashed and dotted lines.} \label{SFR}
\end{figure}

In this section, we use the same method as Y\"{u}ksel et al. (2008)
to calculate the SFR rate from GRBs. Because only very bright bursts
can be seen from all redshifts, we use the same luminosity cuts as
Kistler et al. (2009), as shown in Figure~\ref{GRBnum}. The number
counts in redshift bins $z=4-5$, $5-6$, $6-7$, $7-8.5$ and $8.5-10$
are $10$, $4$, $2$, $1$ and $1$ respectively. In the redshift bin of
$8.5-10$, there is only one GRB named GRB 090429B with photometric
redshift $z\sim 9.4$, although there is a low-probability tail to
somewhat lower redshifts (Cucchiara et al. 2011). The bin choice of
this work is different with those of Robertson \& Ellis (2012). In
this work, we choose redshift bins uniform in $z$, and also ensure
that the number of GRBs in each bins is equal or larger than one. We
also calculate the SFR using bin choice of Robertson \& Ellis
(2012), and find that the result is a little difference compared
with current bin choice. The GRBs in $z = 1-4$ act as a ``control
group'' to constrain the GRB to SFR conversion, since this redshift
bin has both good SFR measurements and GRB counts. We calculate the
theoretically predicated number of GRBs in this bin as
\begin{eqnarray}
N_{1-4}^{\rm the} & = & \Delta t \frac{\Delta \Omega}{4\pi}
\int_{1}^{4} dz\,  F(z) \, \varepsilon(z) \frac{\dot{\rho} _*(z)}
{\left\langle f_{\rm beam}\right\rangle} \frac{dV_{\rm com}/dz}{1+z} \nonumber \\
& = & A \, \int_{1}^{4} dz\, (1+z)^\delta \, \dot{\rho} _*(z) \,
\frac{dV_{\rm com}/dz}{1+z}\,, \label{N1-4}
\end{eqnarray}
where $A = {\Delta t \, \Delta \Omega \, F_0} / 4\pi {\left\langle
f_{\rm beam} \right\rangle}$ depends on the total observed time of
\emph{Swift}, $\Delta t$, and the angular sky coverage, $\Delta
\Omega$. The theoretical number of GRBs in redshift bin $z_1-z_2$
can be written by
\begin{eqnarray}
N_{z_1-z_2}^{\rm th} & = &  \left\langle \dot{\rho} _*
\right\rangle_{z_1-z_2} A \, \int_{z_1}^{z_2} dz\, (1+z)^\delta \,
\frac{dV_{\rm com}/dz}{1+z}\,, \label{Nz1-z2}
\end{eqnarray}
where $\left\langle \dot{\rho}_* \right\rangle_{z_1-z_2}$ is the
average SFR density in the redshift range $z_1-z_2$. Representing
the predicated numbers, $N_{z_1-z_2}^{\rm th}$ with the observed GRB
counts, $N_{z_1-z_2}^{\rm obs}$, we obtain the SFR in the redshift
range $z_1-z_2$,
\begin{equation}
\left\langle \dot{\rho}_* \right\rangle_{z_1-z_2} =
\frac{N_{z_1-z_2}^{\rm obs}}{N_{1-4}^{\rm obs}} \frac{\int_{1}^{4}
dz\, \frac{dV_{\rm com}/dz}{1+z}(1+z)^\delta \dot{\rho}_*(z)\,
}{\int_{z_1}^{z_2} dz\, \frac{dV_{\rm com}/dz}{1+z}(1+z)^\delta }\,.
\label{zratio}
\end{equation}
In the calculation, we assume that the value of $\delta$ is constant
at all redshift range. The derived SFR from GRBs are shown as filled
circles in Figure~\ref{SFR}. Error bars correspond to 68\% Poisson
confidence intervals for the binned events (Gehrels 1986). The
high-redshift SFRs obviously decrease with increasing redshifts,
although an oscillation may exist. We find that the SFR at $z>4.48$
is proportional to $(1+z)^{-3}$ using minimum $\chi^2$ method, which
is shown as solid line in Figure~\ref{SFR}. Because we use different
cosmological parameters comparing to Hopkins \& Beacom (2006)
($\Omega_m=0.3$, $\Omega_\Lambda=0.7$ and $H_0$=70
km~s$^{-1}$~Mpc$^{-1}$), SFR conversion between different cosmology
models must be considered. The conversion factor for a given
redshift range can be expressed as (Hopkins 2004)
\begin{equation}
\dot{\rho}_*(z)\propto \frac{D_{\rm com}^2(z)}{D_{\rm
com}^3(z+\Delta z)-D_{\rm com}^3(z-\Delta z)},
\end{equation}
where $D_{\rm com}$ is given in equation~(\ref{com}). At the
redshift range $z=4-5$, the value of conversion factors in these two
cosmological models are very similar. The relative error is less
than 4\%. So our results are insensitive to the choice of WMAP7
cosmology. The new determination of SFR is slight smaller than the
result given by Kistler et al. (2009). There are two reasons for
this situation. First, we derive a smaller evolution factor index
$\delta$. Second, we update the \emph{Swift} GRB sample. In past
three years, \emph{Swift} has observed much more GRBs with medium
redshifts than GRBs with high redshifts. So the ratio
${N_{z_1-z_2}^{\rm obs}}/{N_{1-4}^{\rm obs}}$ is smaller compared
with Kistler et al. (2009).

Ishida et al. (2011) used the principal component analysis method to
measure the high-redshift SFR from the distribution of GRBs and
found that the SFR at $z\sim 9.4$ could be up to
$0.01M_\odot$~yr$^{-1}$~Mpc$^{-3}$. Robertson \& Ellis (2012)
constrained the SFR using GRBs by considering the contribution of
``dark'' GRBs. They found that the high-redshift SFR derived from
GRBs can vary a factor of 4 using different values of $\delta$.
Their results for $\delta=0.5$ are shown as open circles in
Figure~\ref{SFR}. Our results can be marginally consistent with the
open dots. Elliott et al. (2012) chose 43 GRBs by selecting GRBs
that have been detected by GROND within 4 hours after the Swift BAT
trigger and that exhibited an X-ray afterglow. They found that the
linear relationship between GRB rate and SFR using this small
sample. Johnson et al. (2013) used high-resolution cosmological
simulations to study the high-redshift SFR. Our result is consistent
with that of Johnson et al. (2013) at $z\leq 10$. But at $z\geq 10$,
they found the SFR is reduced by up to an order of magnitude due to
the molecule-dissociating stellar radiation.

\section{GRBs from rapidly rotating metal-poor stars and influence on SFR}

The collapsar model explains GRB formation via the collapse of a
rapidly rotation massive iron core into a black hole (Woosley 1993).
This collapse model requires the initial mass of the massive stars
with masses larger than about $30M_{\odot}$ (Woosley 1993; Bissaldi
et al. 2007). Yoon \& Langer (2005) and Woosley \& Heger (2006)
showed that at low metallicity, quasi-chemically-homogeneous
evolution of rapidly rotating stars with low masses can lead to the
formation of rapidly rotating helium stars which satisfy all the
requirements of the collapsar scenario. Because the rotation affects
the evolution of stars significantly, especially through
rotationally induced chemical mixing (Maeder \& Meynet 2000; Heger
et al. 2000). The star remains chemically homogeneous evolution
scenario (CHES). The CHES is recognized as a promising path towards
collapsars in connection with long GRBs. Yoon, Langer \& Norman
(2006) showed that at low metallicity ($Z\leq 0.004$),
quasi-chemically-homogeneous evolution of rapidly rotating stars
with masses larger than 12$M_\odot$ can lead to long
GRBs\footnote{Although the lower limit mass of a star with low
metallicity that can collapse to GRB is uncertain. But this value is
unimportant in our analysis. The best fitting parameters will shift
slightly when the lower limit mass is changed.}. We call this type
of GRBs as chemically homogeneous GRBs (CHG) below. If stars in the
same mass range have high metallicities and slow rotation, they will
die as type II supernovae (see Figure 3 of Yoon et al. 2006). This
picture has also been confirmed by observation (Fruchter et al
2006).

\begin{figure}
\centering
\includegraphics[width=0.5\textwidth]{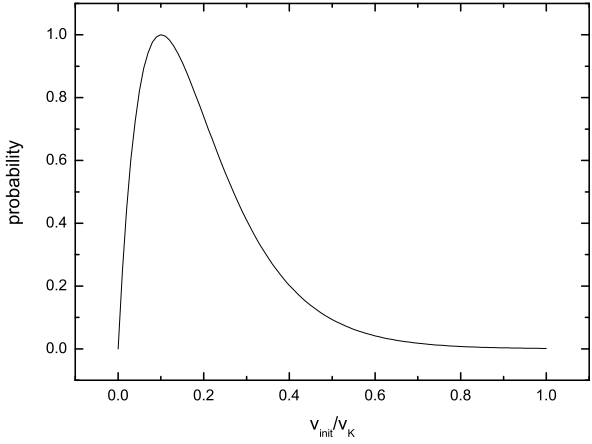} \caption{Distribution
of the initial rotation value of stars.} \label{rotation}
\end{figure}

We study the rate of CHG from chemically homogeneous evolution
scenario as follows. The most widely used functional form for the
initial mass function (IMF) is that proposed by Salpeter (1955):
\begin{equation}
\phi(M)=A_{\rm Salpeter}M^{-2.35},
\end{equation}
where $A_{\rm Salpeter}=0.06$ is the normalization constant derived
from,
\begin{equation}
\int_{m_{\rm low}}^{m_{\rm up}}\phi(M)dM=1.
\end{equation}
We use $m_{\rm low}=0.1M_\odot$ and $m_{\rm up}=120M_\odot$. We
consider stars with masses between $12M_\odot$ and $30M_\odot$.
Because the stars with masses $M\geq30M_\odot$ can produce GRBs
through conventional collapse model (Woosley 1993). So the rate of
CHG is
\begin{eqnarray}
R_{\rm CHG} &=& k_{\rm CHG}\Sigma (Z_{\rm
th},z)\dot{\rho}_*(z)P(x>x_{cr})\int_{12M_\odot}^{30M_\odot}\phi(M)dM,
\end{eqnarray}
where $k_{\rm CHG}$ with value about $10^{-5}$ is the CHGs formation
efficiency, $\Sigma (Z_{\rm th},z)$ and $P(x>x_{cr})$ are discussed
below. According to Langer \& Norman (2006), the fractional mass
density belonging to metallicity below a given threshold $Z_{\rm
th}$ is
\begin{equation}
\Sigma (Z_{\rm th},z)=\frac{\hat{\Gamma}[\alpha_1+2,(Z_{\rm
th}/Z_{\odot})^2 10^{0.15\beta z}]}{\Gamma(\alpha_1+2)},
\end{equation}
where $\hat{\Gamma}$ and $\Gamma$ are the incomplete and complete
gamma functions, $\alpha_1=-1.16$ is the power-law index in the
Schechter distribution function of galaxy stellar masses (Panter,
Heavens \& Jimenez 2004) and $\beta=2$ is the slope of the galaxy
stellar mass-metallicity relation (Savaglio et al. 2005; Langer \&
Norman 2006). We extrapolate the metallicity evolution up to
redshift $z\sim 9.4$. Observation shows that this metallicity
evolution cab be used up to $z\sim 3$ (Kewley \& Kobulnicky 2007).
This extrapolation has been widely used in literature (Langer \&
Norman 2006; Li 2008; Robertson \& Ellis 2012). We set $Z_{\rm
th}=0.004$. As discussed by Yoon, Langer \& Norman (2006), the GRB
production in CHES is limited to metallicity $Z_{\rm th}\leq 0.004$.
Within the CHES, the fraction of stars which forms a long GRB
depends on the semi-convective mixing, and the distribution function
of initial stellar rotation velocities, $D(v_{\rm init}/v_K)$, where
$v_{\rm init}$ is the initial rotation velocity and $v_K$ is the
Keplerian velocity. We use the $v_{\rm init}/v_K$ distribution from
Yoon, Langer \& Norman (2006)
\begin{equation}
D(x)=B\frac{\lambda^\nu}{\Gamma(\nu)}x^{\nu-1}\exp(-\lambda x),
\label{gdis}
\end{equation}
where $\lambda=9.95$, $\nu=2$ and $x\equiv v_{\rm init}/v_K$. The
normalization constant $B=3.2\times 10^5$ is derived from
$\int_0^\infty D(x)dx=1$. They found that this distribution can fit
well with the observational data from Mokiem et al. (2006). Figure
\ref{rotation} shows the numerical value of equation (\ref{gdis}).
In order to produce GRBs, the value of $x$ should be larger than
$0.4$ (Yoon, Langer \& Norman 2006), so
$P(x>x_{cr})=\int_{0.4}^\infty D(x)dx=0.09$.

So the expected number of CHG between redshifts $z$ and $z+\delta z$
can be calculated as
\begin{equation}
N_{\rm CHG}^{\rm exp}(>L)=F\Delta t \frac{\Delta
\Omega}{4\pi}\int_{z}^{z+\delta z} dz \int_{L}^{\infty}dL \Phi(L)
R_{\rm CHG}\frac{dV_{\rm com}/dz}{1+z}, \label{CHG}
\end{equation}
where $\Phi(L)$ is the luminosity  function of GRBs. We use the
Schechter-function form
\begin{equation}
\Phi (L) = \frac{1}{{L_ \star  }}\left( {\frac{L}{{L_ \star  }}}
\right)^\beta  \exp ( - L/L_ \star  ),
\end{equation}
where $\beta=-1.12$ and $L_\star=9\times 10^{52}$erg~s$^{-1}$ (Wang
\& Dai 2011). The integral $\int_{L}^{\infty}dL \Phi(L)$ equals to
$\Gamma(1+\beta, \frac{L}{L_\star})$, where $\Gamma$ is the
incomplete gamma function. Because $1+\beta \rightarrow 0$, we can
approximate $\Gamma(1+\beta, \frac{L}{L_\star}) \rightarrow
-(\frac{L}{L_\star})^{1+\beta}/(1+\beta)$.

We define an effective SFR  $\dot{\rho}_{*}^{\rm eff}$, due to the
CHG, as
\begin{eqnarray}
{ N^{\rm exp}_{\rm CHG}(>L_{ })\over  N_{1-4}^{\rm obs}(>L_{
})}={\int_{z}^{z+\delta z} \varepsilon \dot{\rho}_{*}^{\rm eff}
dV_{\rm com}(z')/(1+z') \over \int_{1}^{4}\varepsilon
\dot{\rho}_*dV_{\rm com}(z')/(1+z')}\label{rateeqn}.
\end{eqnarray}
We consider the star formation history from Hopkins \& Beacom
(2006),
\begin{equation}
\dot{\rho}_*(z)\propto\left\{
\begin{array}{ll}
(1+z)^{3.44},&z<0.97,\\
(1+z)^{-0.26},&0.97<z<4.48,\\
(1+z)^{-7.8},&4.48<z.
\end{array}\right.\label{sfr1}
\end{equation}
with $\dot{\rho}_*(0)=0.02~{\rm M_{\odot}yr^{-1}Mpc^{-3}}$.  For
convenience, here we fit the data by  $N_{1-4}^{\rm obs}(>L_{
})\sim60 L_{ 52}^{-\alpha}$ with $\alpha\sim0.50$. Substituting
Eq.~(\ref{CHG}) into Eq.~(\ref{rateeqn}), we can obtain the
effective SFR as
\begin{eqnarray}
\dot{\rho}_*^{\rm eff}&=&{ F_0\Delta t (\frac{L_{\rm
th}}{L_\star})^{1+\beta}\int_{1}^{4}\dot{\rho}_*(1+z)^{\delta-1}
{dV_{\rm com}}\over (-1-\beta) (1+z)^{\delta} N_{1-4}^{\rm obs}(>L_{
\rm th})}{\Delta \Omega\over
4\pi}{{R_{\rm CHG}(>L_{ \rm th})}}\nonumber\\
&=&C ~{\rm
M_{\odot}yr^{-1}Mpc^{-3}}\int_{12M_\odot}^{30M_\odot}\phi(M)dM~\rho_*(z)(1+z)_1^{1+\alpha-\delta+\beta}
\nonumber\\
&&\times\left({(1+z)^{1/2}-1\over2}\right)^{2(\alpha+\beta+1)},\label{effsfr}
\end{eqnarray}
where the factor $C \sim 125~F_0k_{\rm CHG,-5}F_{\rm
lim,-8}^{1+\alpha+\beta}$. The luminosity threshold at redshift $z$
can be calculated as $L_{ \rm th}=4\pi D_L(z)^2F_{\rm lim}$ for a
given flux sensitivity $F_{\rm lim}$. For the \emph{Swift}
satellite, we adopt the angular sky coverage of $\Delta
\Omega/4\pi\sim 0.1$, and the observation period $\Delta t\sim
7.5$~yr. In equation (\ref{effsfr}), the integral
$\int_{12M_\odot}^{30M_\odot}\phi(M)dM$ is proportional to
$(12M_\odot^{-1.35}-30M_\odot^{-1.35})/1.35=0.026$. Although there
are many factors in equation (19) may subsume the effect in change
the IMF integral for GRB production. The evolution of
$\dot{\rho}_*^{\rm eff}$ is shown as the dotted line in Figure
\ref{SFR}. Bouwens et al. (2009,2011) measured high-redsihft SFR
using color-selected Lyman break galaxies (LBGs) method. Their
results are shown as triangular points in Figure \ref{SFR}. But LBG
studies mainly probe the brightest galaxies. If the integration of
UV luminosity functions down to $M_{UV}\simeq -10$, the SFR inferred
from LBG is consistent with that derived from GRBs (Kistler et al.
2013). GRBs are found to favor sub-luminosity galaxies (Fynbo et al.
2003), so a larger fraction of the SFR within such hosts would be
revealed by GRBs (Kistler et al. 2009). We can see that the SFR
inferred from high-redshift GRBs can be well explained by combining
equation (\ref{sfr1}) with equation (\ref{effsfr}) for $C\sim 125$.
The overall conventional long GRB formation efficiency from massive
stars is about $< 10^{-6}$ (Zitouni et al. 2008; Li 2008), which is
smaller than $k_{\rm CHG}$. This indicates that the subclass of
massive stars with low metallicity and chemical homogeneity may
produce long GRBs more efficiently.

\section{Implications for the cosmic reionization}

\begin{figure}
\centering
\includegraphics[width=0.5\textwidth]{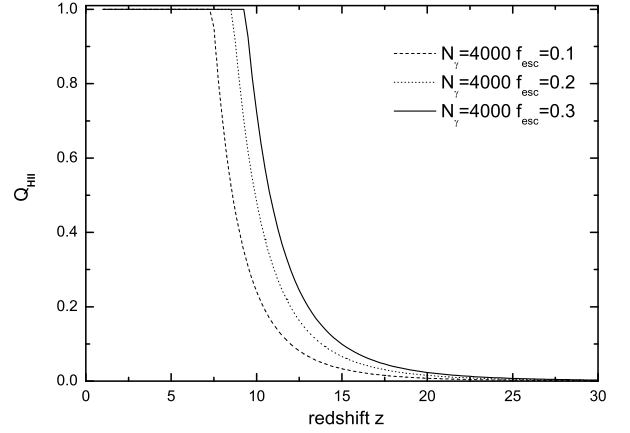} \caption{The HII filling factor $Q_{\rm HII}$ as a function of redshift computed for
different values of $f_{\rm esc}$.} \label{zre}
\end{figure}
Determining when and how the universe was reionized by early sources
have been important questions for decades (Gunn \& Peterson 1965;
Robertson et al. 2010). It is established that intergalactic medium
(IGM) reionization may be completed by $z \approx 6.5$, based on
strong Ly$\alpha$ absorption from neutral hydrogen along lines of
sight to quasars at $z>6$ (Fan et al. 2001). As a measure of
ionization, we follow the evolution of the HII volume filling factor
$Q_{\rm HII}=n_e/n_H$, versus redshift, using the SFR derived from
GRBs (the solid line in Figure~\ref{SFR}). The average evolution of
$Q_{\rm HII}$ is found by numerical integration of the rate of
ionizing photons minus the rate of radiative recombinations (Madau
et al. 1999; Barkana \& Loeb 2001; Wyithe \& Loeb 2003; Yu et al.
2012)
\begin{equation}
    \frac{dQ_{\rm HII}}{dz} = \left( \frac {\dot{N}_{\rm ion}} { n_H } -
           \alpha_B C n_H Q_{\rm HII}\right) \frac{dt}{dz}  \;  .
\label{eq:ionization_history}
\end{equation}
Here,
\begin{equation}
\dot{N}_{\rm ion} =(1+z)^3 \dot{\rho}_*(z)N_\gamma f_{\rm esc}/m_p
\end{equation}
is the rate of ionizing UV photons escaping from the stars into the
IGM, $N_\gamma$ is the number of ionizing UV photons released per
baryon of the stars, $(1+z)^3$ converts the comoving density into
proper density, $\dot{\rho}_*(z)$ is proportional to $(1+z)^{-3}$ at
$z>4.48$ and $f_{\rm esc}$ is the escape fraction. The escape
fraction is not well constrained. At low redshifts, observations
show that the escape fraction from GRB hosts is about a few percent
(Chen et al. 2007; Fynbo et al. 2009). But at high redshifts, the
$f_{\rm esc}$ is larger (Inoue et al. 2005; Robertson et al. 2010).
Recent estimates suggest that the clumping factor $C\approx 1-6$
(Bolton \& Haehnelt 2007; Pawlik et al. 2009). We adopt $C=3$ in
this paper. $n_H$ is the proper density of hydrogen, and
$\alpha_B=1.63\times 10^{-13}\rm cm^{3}~s^{-1}$ is the recombination
rate for an electron temperature of about $10^4$K. Because the mass
in collapsed objects is still small at high redshift, the IGM
contains most of the cosmological baryons, at mean density
\begin{equation}
    \rho_b = \Omega_b \rho_{\rm cr} (1+z)^3 =  4.24 \times 10^{-31}(1+z)^3~{\rm g~cm}^{-3}    \; .
\end{equation}
We adopt the parameters from WMAP seven-year, $\Omega_b h^2 =
0.02255\pm0.00054$ and $\Omega_m h^2 = 0.1352\pm0.0036$ (Komatsu et
al. 2011). The critical density is $\rho_{\rm cr} = 1.8785 \times
10^{-29}~h^2$ g~cm$^{-3}$. The mean hydrogen number density,
\begin{equation}
     n_H = \frac {\rho_b (1-Y)}{m_H} = 1.905 \times 10^{-7}(1+z)^3~{\rm cm}^{-3}       \;
     ,
\end{equation}
where $Y = 0.2477\pm0.0029$ is the helium mass fraction (Peimbert et
al. 2007). After the values of $N_\gamma$ and $f_{\rm esc}$ are
given, the evolution of the HII volume filling factor $Q_{\rm HII}$
can be numerically calculated from equation
(\ref{eq:ionization_history}). In Figure \ref{zre}, we show the
evolution of $Q_{\rm HII}$ as a function of redshift. For
$N_\gamma=4000$ and $f_{\rm esc}=0.2$, the IGM was completely
ionized at $z_{\rm rei}\sim 8.5$.

\begin{figure}
\centering
\includegraphics[width=0.5\textwidth]{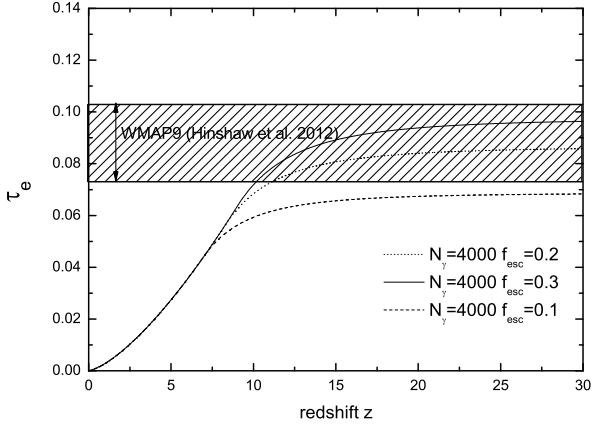} \caption{The optical depth $\tau_e$ due to the
scattering between the ionized gas and the CMB photons is shown. The
shade region is given by the nine-year WMAP measurements
($\tau_e=0.089\pm0.014$). The reionization history calculated from
GRB-inferred SFR can easily reach $\tau_e$ from WMAP nine-year
data.} \label{tau}
\end{figure}

The cosmic microwave background (CMB) optical depth back to redshift
$z$ can be written as the integral of $n_e \sigma_T d \ell$, the
electron density times the Thomson cross section along proper
length,
\begin{equation}
   \tau_e(z) = \int _{0}^{z} n_e(z) \sigma_T (1+z')^{-1} \; [c/H(z')] \; dz'    \;   .
\end{equation}
So after we obtain the redshift evolution of $n_e(z)$, the CMB
optical depth as a function of redshift can be calculated. More
recently, the Planck team has released the latest result on
cosmological parameters (Plack Collaboration 2013). In the
calculation, we extrapolate the SFR as $(1+z)^{-3}$ to $z\sim 30$.
The first stars, so-called Population III (Pop III) stars are
predicated to have formed at $z>20$ in minihalos (Tegmark et al.
1997; Yoshida et al. 2003). Heger et al. (2003) and M\'{e}sz\'{a}ros
\& Rees (2010) show that Pop III stars can die as GRBs. The
formation rate of Pop III GRBs has been extensively studied (Campisi
et al. 2011; de Souza et al. 2011). The high luminosities of GRBs
make them detectable out to the edge of the visible universe (Bromm
\& Loeb 2002, 2006; Wang et al. 2012). So GRBs may provide the
information of SFR out to $z>20$ in future. The extrapolation of SFR
to high redshifts may be reasonable. The optical depth is shown in
Figure \ref{tau}. The WMAP nine-year data gives $\tau_e=0.089\pm
0.014$ (Hinshaw et al. 2012), which is shown as the shaded region.
The combination of Plack and WMAP data also gives
$\tau_e=0.089_{-0.014}^{+0.012}$ (Plack Collaboration 2013). So our
GRB-inferred SFR can reproduce the CMB optical depth.

\section{Summary}
Using the GRB catalogs, we have constructed the cumulative redshift
distribution of 110 luminous ($L_{\rm iso} > 10^{51} \rm
erg~s^{-1}$) GRBs out to redshift $z\sim 9.4$. We find that the
\emph{Swift} GRBs during 2005-2012 are biased toward tracing the
SFR, including a factor of about $(1+z)^{0.5}$. Correcting this
evolution, we derive the star formation history up to $z\sim 9.4$
using \emph{Swift} GRB sample. Our results show that no steep drop
exists in the SFR up to at least $z\sim 9.4$. In order to explain
the high-redshift GRB rate excess, the GRBs produced by rapidly
rotating metal-poor stars with low mass are considered. The
collapsar model explains GRB formation via the collapse of a massive
star with $M>30M_\odot$ into a black hole. We consider that at low
metallicity, quasi-chemically homogeneous evolution of rapidly
rotating stars with mass larger than $12M_\odot$ can lead to the
formation of GRBs. The low metallcity and rapid rotation can lead to
efficiently produce GRBs in two ways. First, rapid rotation keeps
the stars chemically homogeneous and thus avoids the formation of a
massive envelope, so stellar core is free of spin-down due to
magnetic core-envelope coupling. Second, the stellar wind is weak at
low metallicity, so this reduces spin-down due to stellar winds. Our
fitting results confirm this idea. We also calculate the
reionization history using the GRB-inferred SFR, and find that this
SFR can maintain cosmic reionization over $6<z<10$ and reproduce the
observed optical depth of Thomson scattering to the cosmic microwave
background.

\acknowledgements We thank the anonymous referee for very useful
comments and suggestions. We thank K. S. Cheng and Z. G. Dai for
fruitful discussion. We acknowledge the use of public data from the
\emph{Swift} data archive. This work is supported by the National
Natural Science Foundation of China (grant 11103007 and 11033002).

\end{document}